\documentclass[twocolumn]{autart}

\usepackage{graphicx}          
                               
\usepackage{color}
\usepackage[version=3]{mhchem}
\usepackage[dvipsnames]{xcolor}

\usepackage[figuresright]{rotating}
\usepackage{hyperref}
\usepackage{cuted}   
\usepackage{soul,amsmath,amsfonts}
\usepackage{multirow}	
\usepackage{multicol} 
\usepackage{lipsum}

\usepackage[normalem]{ulem} 

\usepackage{wasysym} 

\journal{Physics Letters B}

\def\V{\text{Var}}

\def\RR{R_{4/2}}

\usepackage{subfigure}

\begin{document}

\begin{frontmatter}
\title{Nuclear incompressibility and fourth moment of the nuclear density in Skyrme functionals}

\author[CENS,IRIS]{Soonchul Choi}, 
\ead{scchoi0211@ibs.re.kr}
\author[CENS]{Tae-Sun Park}, 
\ead{tspark@ibs.re.kr}
\author[IRIS]{Panagiota Papakonstantinou}
\ead{ppapakon@ibs.re.kr}

\address[CENS]{Center for Exotic Nuclear Studies, Institute for Basic Science (IBS), Daejeon 34126, Republic of Korea}
\address[IRIS]{Institute for Rare Isotope Science, Institute for Basic Science, Daejeon, 34000, Korea}

\begin{abstract}
Recent experimental advances could soon allow the accurate extraction of not only the root-mean-square 
radius but also the fourth radial moment of the nuclear electric charge density distribution. 
The fourth radial moment of the nuclear density distribution, $R_4\equiv\sqrt[4]{\left<r^4\right>}$, provides a sensitive probe of the nuclear surface thickness, as it is more susceptible to the large-$r$ distributions than the root-mean-square radius ($R_2$). In this work, we examine the utility of $R_4$ for constraining the nuclear equation of state (EoS) at subsaturation densities,  specifically for the proton distribution and within the framework of Skyrme energy density functionals. Using a statistical analysis based on predictions from one hundred Skyrme functional models, we demonstrate strong correlations between the energy per particle curvature $K(\rho)$ at $\rho = 0.08 \text{ fm}^{-3}$ and $R_4$ (or the ratio $R_{4/2}=R_4/R_2$) in representative nuclei such as $\text{}^{48}\text{Ca}$ and $\text{}^{208}\text{Pb}$. 
We establish 
that $R_{4/2}$, being sensitive to the density tail, serves as an efficient proxy for sub-saturation $K(\rho)$ within the tested Skyrme functional space. Knowledge of $R_{4/2}$ within 0.5\% precision or 
better,
for example in $^{48}$Ca or $^{208}$Pb, could constrain the curvature of the energy per particle of symmetric matter at $0.08$ fm$^{-3}$ within 20 MeV or 
less.
  \end{abstract}
  
\end{frontmatter}



\section{Introduction}


Charge distributions and their second moments have been measured accurately in stable nuclei via electron scattering and other electromagnetic probes~\cite{Sick:2014sra,deVries1987}. Recent experimental developments make it possible to extract information also for the weak charge distributions \cite{PREX:2021umo,CREX:2022} or to probe exotic nuclei~\cite{Suda2025}. 
Related theoretical work focuses especially on the possibility to determine the neutron skin thickness more accurately than has been possible so far~\cite{Liu:2021ofe,Reinhard:2019ixi,Naito:2021gre,Xie2024,Miyagi2025}. 
A relationship between the charge form factor and the moments of the proton and neutron distributions has been explored in \cite{Kurasawa2020}.

In this work, we first demonstrate that the extension of the surface region in stable nuclei is largely determined by the curvature of the energy per particle in symmetric nuclear matter $K(\rho)$ at sub-saturation densities. Being higher order, the fourth moment of the proton, neutron, or charge density distribution shows greater sensitivity to the surface region than the second moment. 
The surface thickness, in turn, is related to the moments of the density, 
as can be demonstrated using analytical models such as the Helm model~\cite{ANDRAE2000413}.
It is thus natural to ask whether information on the nuclear equation of state (EoS) at sub-saturation densities can be extracted from measurements of both the second and the fourth moments of the charge density distribution or, equivalently, from both the size and the surface thickness of a nucleus. 


%
%
%
The connection of the nuclear incompressibility with the density profile is not a new idea. 
The relation of the nuclear compression modulus with 
the surface tension and the density distribution near the surface region is known in the droplet model and has been explored microscopically in semi-infinite nuclear matter~\cite{Blaizot1980,DCV1999}. 
The present work investigates whether an analogous connection can be quantified in finite nuclei through the measurable quantity $\RR\equiv R_4/R_2 $, which is chosen as a measure of the surface thickness for a given $R_2$.
We study the correlation
by using
the results of one hundred Skyrme models \cite{Dutra:2012mb}. 



The organization of the paper is as follows: 
Section \ref{sec:formalism} presents the formalism for the microscopic Skyrme model, the Helm model of nuclear densities, the EoS for nuclear matter, and the quantification of linear correlations involving two or three quantities. 
Section \ref{sec:compr} expands on the motivation for this study by demonstrating the relationship between $K(\rho)$ and fourth moment using numerical results. 
In Section \ref{sec:mpr}, we address the question of constraining $K(\rho)$ from measurements of the density moments. 
We conclude in Section \ref{sec:summary}.


\section{Formalism\label{sec:formalism}}

\subsection{Skyrme energy density functional}
The present study requires a large set of realistic  
nuclear models
that describe 
the charge density distribution $\rho_{\mathrm ch}$, 
the resulting $R_2$ and $R_4$,
as well as $K(\rho)$ and other EoS parameters microscopically. 
To this end, we employ primarily the Skyrme energy density functional,
which is widely used in nuclear physics 
due to its 
computationally efficient framework for describing both finite nuclei and infinite nuclear matter on an equal footing. It is based on a density-dependent effective interaction that accounts for the interplay between individual nucleons and the surrounding nuclear medium.
The Skyrme interaction between two nucleons at coordinates 
$\boldsymbol{r}_1,\boldsymbol{r}_2$  is written as:
 \begin{eqnarray}
    V_{12}(\boldsymbol{r}_1,\boldsymbol{r}_2)&=&t_0(1+x_0\boldsymbol{P}_\sigma)\delta(\boldsymbol{r})
    \nonumber\\
    &+&\frac{1}{6}t_3(1+x_3P_\sigma)\rho^\sigma(\boldsymbol{R})\delta(\boldsymbol{r})
    \nonumber\\
    &+&\frac{1}{2}t_1(1+x_1P_\sigma)(\boldsymbol{K}'^2\delta(\boldsymbol{r})+\delta(\boldsymbol{r})\boldsymbol{K}^2)
    \nonumber\\
    &+&t_2(1+x_2P_\sigma)\boldsymbol{K}'\cdot\delta(\boldsymbol{r})\boldsymbol{K}
    \nonumber\\
    &+&iW_0(\sigma_1+\sigma_2)\cdot[\boldsymbol{K}'\times\delta(\boldsymbol{r})\boldsymbol{K}]
    \label{Skyrme_eq}
 \end{eqnarray}
where $\boldsymbol{r}=\boldsymbol{r}_1-\boldsymbol{r}_2$,  $\boldsymbol{R}=(\boldsymbol{r}_1+\boldsymbol{r}_2)/2$,
$P_\sigma$ is the spin exchange operator and
$\sigma_i$ are the Pauli spin matrices.
In the above equation, the relative momentum operator $\boldsymbol{K}=(\nabla_1-\nabla_2)/2i$ and $\boldsymbol{K}'=-(\nabla_1-\nabla_2)/2i$ act on the wave function to the right and left, respectively. 

The energy per particle $E(\rho,\delta)$ in infinite nuclear matter with a given baryon density $\rho=\rho_p+\rho_n$ and isospin asymmetry $\delta=(\rho_n-\rho_p)/\rho$, where $\rho_n$ and $\rho_p$ denote the neutron and proton density, respectively, can be written analytically as follows:
\begin{eqnarray}
    E(\rho,\delta) &=& \frac{\hbar C}{2}\rho^{2/3}
    \left(\frac{1}{M_n}\left(\frac{1-\delta}{2}\right)^{5/3}
    +\frac{1}{M_p}\left(\frac{1+\delta}{2}\right)^{5/3}\right)
    \nonumber\\
    &+& C\rho^{5/3}\left[\left(\frac{1-\delta}{2}\right)^{5/3} + \left(\frac{1+\delta}{2}\right)^{5/3}\right]
    \nonumber\\
    &&\times\left[\frac{t_1}{4}\left(1+\frac{x_1}{2}\right) + \frac{t_2}{4}\left(1+\frac{x_2}{2}\right)\right]
    \nonumber\\
    &+&C\rho^{5/3}\left[\left(\frac{1-\delta}{2}\right)^{8/3} + \left(\frac{1+\delta}{2}\right)^{8/3}\right]
    \nonumber\\
    &&\times\left[\frac{t_2}{4}\left(\frac{1}{2}+x_2\right) + \frac{t_1}{4}\left(\frac{1}{2}+x_1\right)\right]
    \nonumber\\
    &+&\frac{t_0}{2}\left[\left(1+\frac{x_0}{2}\right)-\left(\frac{1}{2}+x_0\right)\left(\frac{1+\delta^2}{2}\right)\right]\rho
    \nonumber\\
    &+&\frac{t_3}{12}\left[\left(1+\frac{x_3}{2}\right)-\left(\frac{1}{2}+x_3\right)\left(\frac{1+\delta^2}{2}\right)\right]\rho^{\sigma+1},
\end{eqnarray}
where $C=\frac{3}{5}(3\pi^2)^{2/3}$.  
The properties of finite nuclei are calculated with the Skyrme-Hartree-Fock (SHF) approximation, for which we use the publicly available code of ref.~\cite{Reinhard:2021yke}.
The moments of the densities are calculated from the numerically obtained density profiles.

To study the correlation between the charge fourth radial moment and nuclear matter properties, we use the following
100 representative Skyrme interactions \cite{Dutra:2012mb}: BSk1, BSk10, BSk11, BSk12, BSk13, BSk14, BSk15, BSk16, BSk17, BSk2, BSk2p, BSk3, BSk4, BSk5, BSk6, BSk8, BSk9, E, Es, Gs, KDE, KDE0v, KDE0v1, MSK1, MSK2, MSK3, MSK4, MSK5, MSK5p, MSK6, MSK7, MSK8, MSK9, MSL0, MSL1, NRAPR, RATP,  Rs, SAMi, SG1, SG2, SK255, SK272, SKa, SKb, SkI3, SkI4, SkI5, SKM, SKMP, SKMs, SKP, SkS1, SkS2, SkS3, SkS4, SkSC15, SKSC4, SkT1, SkT1s, SkT2, SkT3, SkT3s, SkT4, SkT5, SkT6, SkT7, SkT8, SkT9, SKX, SKXce, SKXm, SKxs15, SKxs20, SKxs25, Skz1, Skz2, Skz3, Skz4, SLy10, SLy230a, SLy4, SLy5, SLy6, SLy7, SQMC600, SQMC650, SQMC700, SVbas, SVmin, v070, v075, v080, v090, v100, v105, v110, Z, Zs, Zss.
%
To discuss the model dependence of our results, we also employ a few relativistic mean field models, see Sec.~\ref{sec:mpr}.
The charge distributions are obtained from the point-proton and point-neutron densities using the charge form-factor prescription implemented in the SKYAX Skyrme-Hartree-Fock code~\cite{Maruhn2005}. Specifically, the charge form factor is written as
\begin{equation}
  F_{\rm ch}(q)
  =
  F_p(q)G_E^p(q)+F_n(q)G_E^n(q),
\end{equation}
where \(F_p(q)\) and \(F_n(q)\) are the Fourier-Bessel transforms of the point-proton and point-neutron densities, respectively, and \(G_E^p(q)\) and \(G_E^n(q)\) are the standard proton and neutron electric Sachs form factors. The charge density is then obtained by the inverse Fourier-Bessel transform,
\begin{equation}
  \rho_{\rm ch}(r)
  =
  \frac{1}{2\pi^2}
  \int_0^\infty q^2 F_{\rm ch}(q) j_0(qr)\,dq .
\end{equation}
The charge-density moments used in this work are calculated as
\begin{equation}
R_n =
  \left[
  \frac{\int_0^\infty 4\pi r^{2+n}\rho_{\rm ch}(r)\,dr}
  {\int_0^\infty 4\pi r^2\rho_{\rm ch}(r)\,dr}
  \right]^{1/n},
\end{equation}
$n=2,\,4$.  
We note that the root-mean-square of the charge distribution is quite precisely known for many stable nuclei. 
For example, for $^{40}\textmd{Ca}$, the $R_2$ is reported as $4.4776\pm0.0019$ fm \cite{Angeli:2013epw}, {\em i.e.,} with an uncertainty of $0.04\%$. 
Experimentally known $R_2$ values are often used in the fitting protocols of Skyrme functionals.
A comparable precision for the fourth moment of the charge density is not available at present. For example, values extracted for the surface thickness of many nuclei from analyses of electron-scattering form factors are found to be of the order of 1~fm and have been reported to the third decimal place~\cite{Friedrich:1982esq}, which implies uncertainties of the order of 0.1\%. Therefore, for simplicity, and for the purposes of this work, we will consider the $R_2$ as precisely known and focus on the uncertainties in $R_4$.


\subsection{$\RR$ and surface thickness} 
\label{RR_and_surface}
The nuclear surface is diffuse, as is known from electron scattering experiments~\cite{Friedrich:1982esq}. As a measure of the diffuseness, or the surface thickness, we adopt $\RR$. 
Before analyzing the microscopic results obtained with the Skyrme functionals, it is useful to establish the relationship between the surface thickness and $\RR$ with a simple analytical model.

Many analytical models of the density profile are  available~\cite{Myers:1983seb,Papakonstantinou:2013rda,ANDRAE2000413}.
One of them is the Helm model, which is obtained by a convolution 
\begin{equation}
\rho(r) = Z \int d^3r' \rho_H(r') \rho_G (|\vec{r}-\vec{r'}|) 
\end{equation}
of a homogeneous sphere of radius $R_H$ representing the bulk of the nucleus, $\rho_H(r)=3\Theta(R_H-r)/(4\pi R_H^3)$, 
and a Gaussian distribution of width $R_G$ representing the surface thickness, 
$\rho_G(r)=\exp{(-r^2/R_G^2)}/(\pi^{3/2}R_G^3)$.
The model can thus accommodate profiles of light nuclei which comprise mostly of a surface part as well as of heavy nuclei with a saturated core. 
In the Helm model, the 
ratio $R_4/R_2$ is given as
\begin{equation}
    R_{4/2}\equiv 
    \frac{R_4}{R_2}=\frac{\left[\frac{15}{4}(1+\frac{4}{5}u^2+\frac{4}{35}u^4)\right]^{1/4}}{\left[\frac{3}{2}(1+\frac{2}{5}u^2)\right]^{1/2}}
\end{equation}
where $u=R_H/R_G$. 
$\RR$ decreases monotonically with $u$. 
Thus, for a given nucleus, the ratio $\RR$ is smaller when the surface is thinner (larger $u$) and larger when the surface is thicker (smaller $u$).
Therefore, $\RR$ represents a measure of the surface thickness for a given nucleus. 

For vanishing surface thickness ($u \to \infty$), we recover the ratio for a homogeneous sphere, $R_4/R_2 \to (25/21)^{1/4} \approx 1.04455$.
In the opposite limit, where the surface thickness is large compared to the radius ($u \to 0$), the ratio approaches that of a Gaussian distribution, $R_4/R_2 \to (5/3)^{1/4} \approx 1.1362$.

\subsection{Equation of state\label{sec:eos}}
The EoS for isospin asymmetric matter can be expressed as a power series of the energy per particle to second order in isospin asymmetry $\delta=(\rho_n-\rho_p)/\rho$,
\begin{equation}
    E(\rho,\delta)=E_0(\rho)+E_{\textmd{sym}}(\rho)\delta^2+O(\delta^4),
\end{equation}
where $E_0(\rho) =E(\rho,\delta=0)$ is the binding energy per nucleon of symmetric nuclear matter, 
and
\begin{equation}
    E_{\textmd{sym}}(\rho)=\left. \frac{1}
    {2}\frac{\partial^2 E(\rho,\delta)}{\partial\delta^2}\right|_{\delta=0},
\end{equation}
is the symmetry energy.
$E_0(\rho)$ and 
$E_{\textmd{sym}}(\rho)$ can be expanded around the nuclear matter saturation density, $\rho_0$,
\begin{eqnarray}
   E_0(\rho) &=& E_0(\rho_0) + \frac{K_0}{2}\chi^2 + O(\chi^3)
   \nonumber\\
   E_{\textmd{sym}}(\rho) &=& E_{\textmd{sym}}(\rho_0) + L\chi + \frac{K_{\textmd{sym}}}{2}\chi^2  + O(\chi^3),
\end{eqnarray}
where $\chi=(\rho-\rho_0)/3\rho_0$.
The coefficient of each term is defined by 
\begin{equation} 
K_0=K(\rho_0) \,\, , \,\, L = L(\rho_0) \,\, , \,\, K_{\textmd{sym}} = K_{\textmd{sym}}(\rho_0), 
\end{equation}
where we have defined the density-dependent parameters 
\begin{eqnarray}
    K(\rho)&=&9\rho^2\frac{\partial^2E_0(\rho)}{\partial \rho^2},\quad \\
    L(\rho)&=&3\rho\frac{\partial E_{\textmd{sym}}(\rho)}{\partial \rho},\quad\\
    K_{\textmd{sym}}(\rho)&=&9\rho^2\frac{\partial^2E_{\textmd{sym}}(\rho)}{\partial \rho^2}.
\end{eqnarray}
For a Skyrme functional, in particular, the curvature of the energy per particle in symmetric nuclear matter is given by
\begin{eqnarray}
    K(\rho) &=& -\frac{\hbar^2 C}{2^{5/3}}\left(\frac{1}{M_n}+\frac{1}{M_p}\right)\rho^{2/3}
    \nonumber\\
    &+& 5(2)^{1/3} C\rho^{5/3} \left[\frac{t_1}{4}\left(1+\frac{x_1}{2}\right) + \frac{t_2}{4}\left(1+\frac{x_2}{2}\right)\right]
    \nonumber\\
    &+&5(2)^{-2/3}C\rho^{5/3} \left[\frac{t_2}{4}\left(\frac{1}{2}+x_2\right) - \frac{t_1}{4}\left(\frac{1}{2}+x_1\right)\right]
    \nonumber\\
    &+&9\frac{t_3}{24}\left[(2+x_3)-(\frac{1}{2}+x_3)\right]\sigma(\sigma+1)\rho^{\sigma+1},
\end{eqnarray}
where 
$C=\frac{3}{5}(3\pi^2)^{2/3}$.  The nuclear compression modulus at saturation density, $K_0\equiv K(\rho_0)$,
is reported to lie
between $200-250$MeV~\cite{Dutra:2012mb,Roca-Maza:2018ujj}. 

The premise of the present work is that the curvature of symmetric matter $K(\rho)$ is the most relevant one for studying the moments of the density. We therefore proceed to substantiate its role.  

\section{Relevance of $K(\rho$) \label{sec:compr}}

First, we present an intuitive argument for the relevance of $K(\rho)$ in studying the moments of the density supported by numerical examples. 
Next, with the help of a correlation analysis, we determine that $K(\rho)$ at approximately 
$0.08$~fm$^{-3}$ is most related with the density profile in nuclei with moderate neutron excess.
We also compare correlations of all the low-order EoS parameters with the moments of the density within the Skyrme model.

\subsection{Demonstration with representative models}

The diffuseness of the nuclear surface is intuitively expected to be sensitive to the compressibility of nuclear matter:
A stiffer EoS penalizes departures from the equilibrium density more strongly and therefore disfavors the extension of matter into the low-density surface region. Conversely, a softer EoS allows a more diffuse surface. Since $\RR$ is particularly sensitive to the density tail and therefore the surface diffuseness, one expects it to contain information on the curvature of the EoS at subsaturation densities.

The above intuition is not new and has been borne out by microscopic calculations for semi-infinite matter \cite{Blaizot1980,DCV1999}. 
For example, in Ref.~\cite{DCV1999}, the surface properties of semi-infinite nuclear matter were studied within relativistic effective field theory and for three values of the compression modulus $K_0$ equal to $125, 200$, and $350$ MeV. 
A number of physical parameters were found to affect the surface thickness but when all else is equal, a higher compression modulus results in a lower surface thickness. In the droplet model, incompressibility affects also the central density: a higher $K_0$ can lead to a slightly increased interior density, and thus a smaller overall nuclear size~\cite{Myers:1983seb}. 

\begin{figure}[h]
    \begin{center}
    \includegraphics[width=7cm]{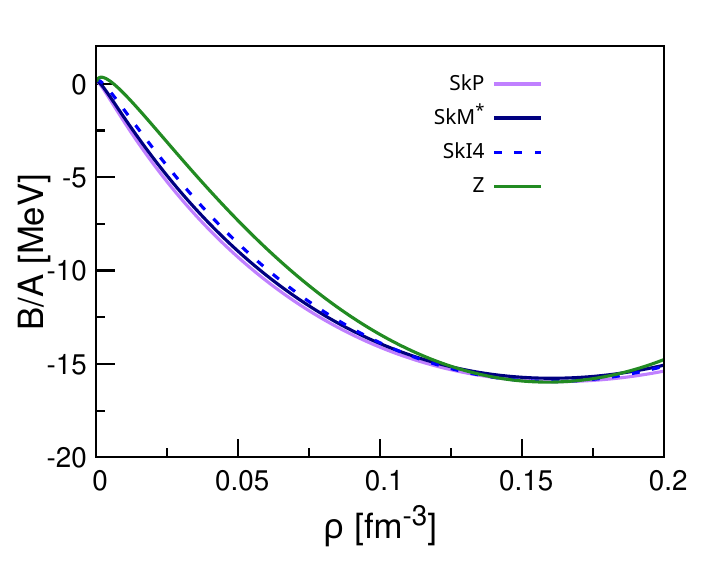}
    \includegraphics[width=7cm]{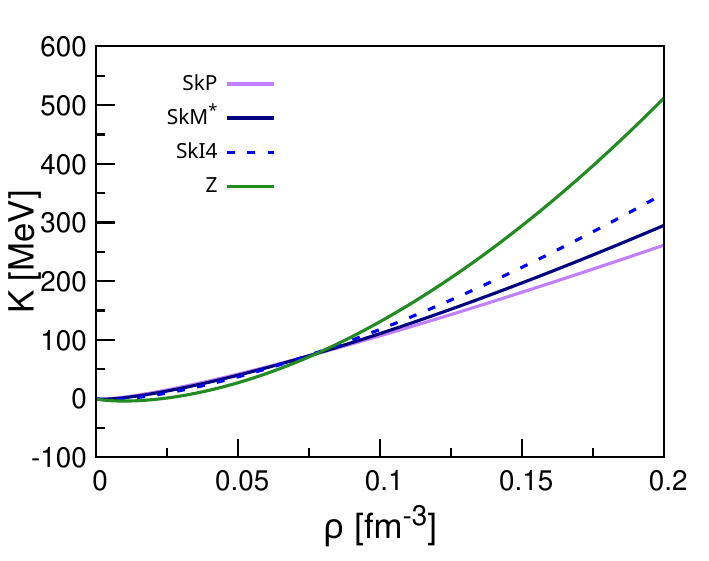}
    \end{center}
    \caption{Energy per particle $E(\rho)$ and curvature $K(\rho)$ as a function of the density in the symmetric nuclear matter for four representative Skyrme functionals}
     \label{fig:1}
\end{figure}

Unlike neutron-skin observables, for which droplet-model expressions can be derived analytically, we are not aware of a comparable macroscopic relation linking $\RR$ to $K (\rho)$. Therefore the quantitative relation is investigated microscopically, with the help of the SHF model.
Among the set of Skyrme functionals considered in this study, 
we select four representative models that span a range of incompressibility values
$K_0$ and 
saturation densities, as summarized in Table~\ref{tab:1}. 
Figure~\ref{fig:1} shows the energy per nucleon $E(\rho)$ and the curvature $K(\rho)$ as functions of the density in symmetric nuclear matter for these Skyrme functionals.
As seen in the figure, functionals with higher $K_0$
exhibit larger energies at sub-saturation densities.
Furthermore, the curvature $K(\rho)$  at low densities tends to be smaller for models with larger $K_0$.
However, this trend may be a model-dependent feature of Skyrme interactions, where the skewness parameter
$Q_0$, related to the third derivative of $E(\rho)$,
is constrained by the lower-order terms in the functional.


\begin{table}[h!]
\centering
\begin{tabular}{|l|cccc|}
\hline
 & SKP & SkM* & SkI4 & Z \\
\hline
$K_0$ (MeV) & 200 & 216.04 & 247.65 & 330 \\
$\rho_0$ (fm$^{-3}$) & 0.162 & 0.160 & 0.160 & 0.159 \\
\hline
$^{16}$O & 2.831 & 2.800 & 2.719 & 2.664 \\
$^{40}$Ca & 3.539 & 3.517 & 3.446 & 3.395 \\
$^{56}$Ni & 3.813 & 3.775 & 3.792 & 3.683 \\
$^{100}$Sn & 4.527 & 4.495 & 4.493 & 4.398 \\
\hline
\end{tabular}
\caption{
$K_0$, $\rho_0$, and the charge radii $R_2$ (in units of fm)  for four nuclei, calculated using each of the four selected Skyrme functionals.}
\label{tab:1}
\end{table}

Table~\ref{tab:1} also shows the $R_2$ values obtained for the four selected nuclei.
As expected, 
the radii are found to be larger for the soft functional with lower 
$K_0$, and smaller for the stiff functional.

\begin{figure*}[h!]
    \begin{center}
    \includegraphics[width=7cm]{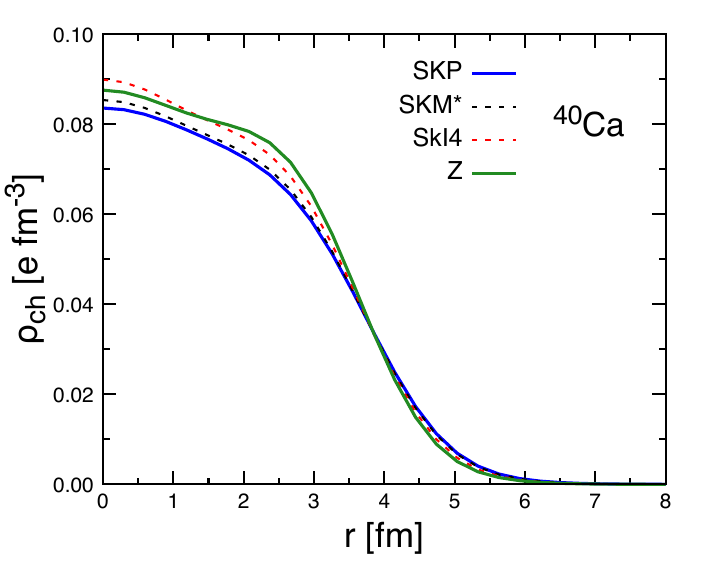}
    \includegraphics[width=7cm]{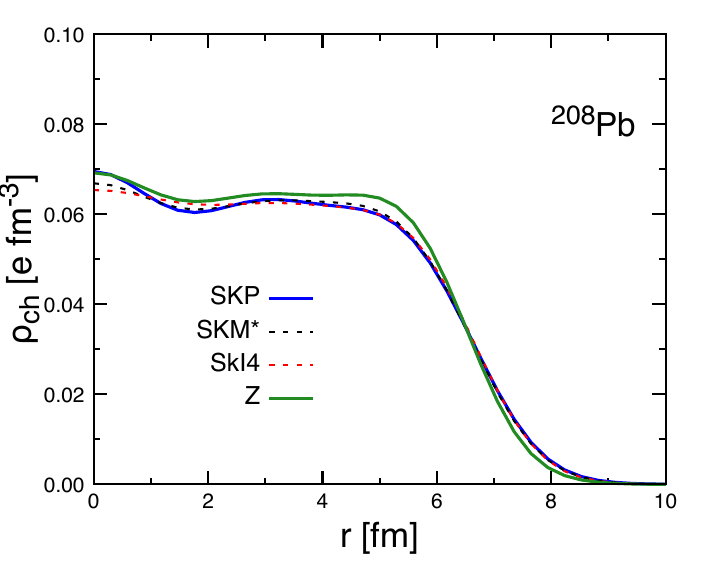}
    \includegraphics[width=7cm]{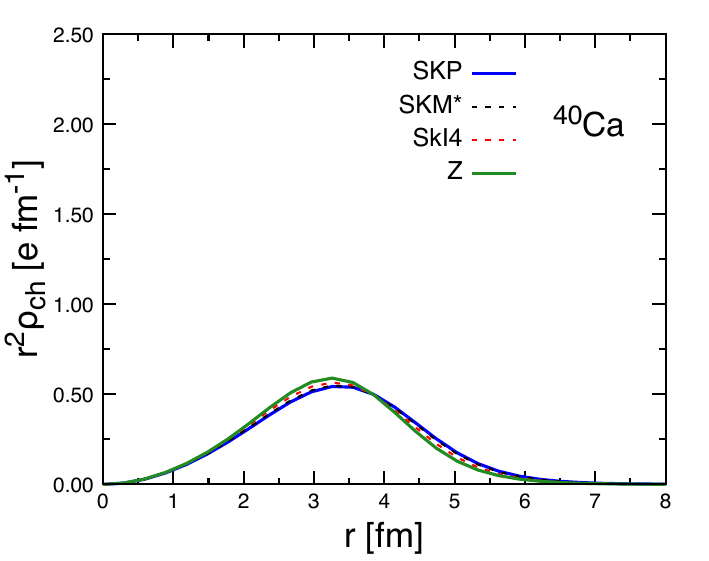}
    \includegraphics[width=7cm]{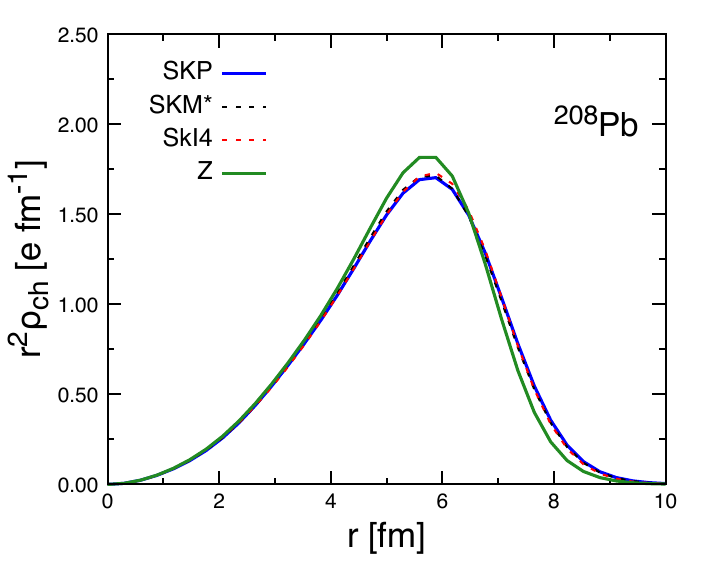}
    \includegraphics[width=7cm]{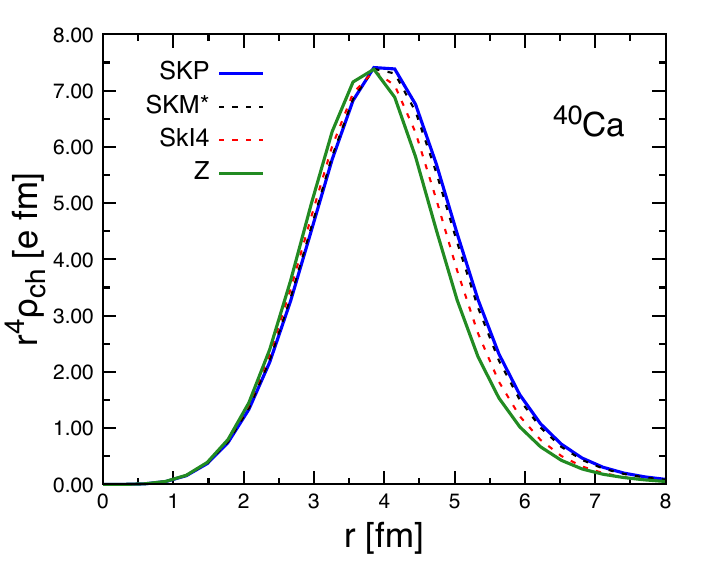}
    \includegraphics[width=7cm]{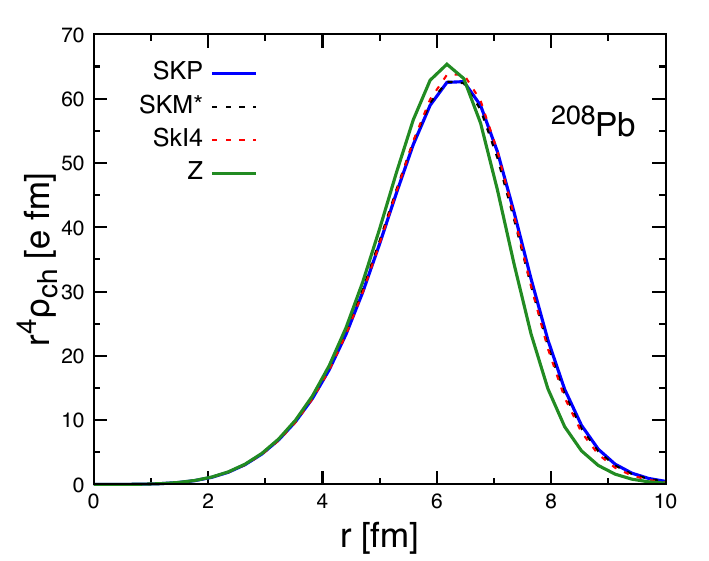}
    \end{center}
    \caption{Proton density profile of $^{40}\textmd{Ca}$ and $^{208}\textmd{Pb}$ obtained with the indicated functionals, which correspond to the equations of state shown in Fig.~\ref{fig:1}. The density profiles are shown also scaled with $r^2$ (middle) and with $r^4$ (bottom)}
     \label{fig:2}
\end{figure*}

In Figure~\ref{fig:2},
we show 
the charge density profiles for $^{40}\textmd{Ca}$ and
$^{208}\textmd{Pb}$ including the profiles weighted by $r^2$ and by $r^4$.
It is evident that models with higher $K_0$ shift the density distribution inward,
and this effect becomes more pronounced as the power $n$ 
in the weighting factor $r^n$ increases.
We observe that the quantity $r^n \rho_p$  extends furthest outward for the softest model considered, SkP, with the effect becoming increasingly pronounced for larger values of $n$. 
We also find that SkP yields the lowest central density, although it corresponds to the highest $\rho_0$ among the models in Table~1. This may not be entirely counterintuitive, as the soft EoS allows nucleons to be more easily extended outward, thereby reducing the central density.

Interestingly, all the considered density profiles intersect at approximately $\rho_p\simeq 0.04\ \text{fm}^{-3}$. 
The crossing of the density profiles near $\rho_{\mathrm{ch}}\approx0.04$~fm$^{-3}$ reflects the fact that the various interactions generate density distributions with different diffuseness but the same normalization $Z$ 
and overall size. 
%
Such profiles naturally intersect near the half-density point,
see Appendix A for a qualitative explanation.
Note that the charge density is primarily determined by the proton distribution, whose bulk density corresponds to half the saturation density so that half its bulk density is approximately a quarter of the saturation density of roughly $0.16$~fm$^{-3}$. 
The crossing point would be different in very neutron-rich nuclei. For example, calculations in $^{60}$Ca (not shown), where the bulk proton density is lower than in stable nuclei, yield a crossing at lower densities. 
Such nuclei have a very neutron-rich surface and are not good candidates to constrain the incompressibility of symmetric matter. 
We therefore consider only nuclei with more moderate neutron excess.

The representative models shown in Fig.~\ref{fig:1}
also exhibit a crossing
at approximately 
half the saturation density, which is the typical total density on the surface of stable nuclei where the density profiles cross as well.

\subsection{Density-dependent correlations}

Having illustrated the premise of this study, we aim to quantify the correlation—if any—between the nuclear matter incompressibility and the density profiles of stable or isospin-symmetric nuclei represented by $R_2$ and $\RR$.  We first examine the strength of these correlations and identify the density region where they are most pronounced. We then investigate whether a highly precise measurement of the fourth moment, combined with an accurate value of the second moment, could place meaningful constraints on  
$K_0$, or more generally, on the density-dependent curvature $K(\rho)$.

For  all 
the Skyrme functionals adopted in this study, 
we have calculated the density-dependent EoS parameters 
$X(\rho)=E(\rho)$, $K(\rho)$, $E_{\textmd{sym}}(\rho)$, $L(\rho)$, $K_{\textmd{sym}}(\rho)$.
%
To explore the correlations of these 
with $R_2$ and $\RR$, we 
employ
the Pearson correlation coefficient (PCC) 
as well as the multiple correlation coefficient (MCC). 

The 
PCC
is a measure of the strength and direction of a linear relationship between two variables:
\begin{equation}
    r_{XY} = 
    \frac{\frac{1}{n-1}\sum^n_i(Y_i-\bar{Y})(X_i-\bar{X})}{\sqrt{\frac{1}{n-1}\sum^n_i(X_i-\bar{X})^2}\sqrt{\frac{1}{n-1}\sum^n_i(Y_i-\bar{Y})^2}},
\end{equation}
which approaches
$+1$ ($-1$) 
in the case of a perfect positive (negative) correlation.
Here, $X$ 
represents one of the EoS parameters at a given density:
$E_0(\rho)$, $E_{\textmd{sym}}(\rho)$, $L(\rho)$, $K(\rho)$, and $K_{\textmd{sym}}(\rho)$,
while $Y$ is either $R_2$ or the ratio $\RR$. 
As defined above, the nuclear matter properties are a function of density, so the correlation coefficient is also a function of density.
Studying the correlation not only at saturation density but also across a range of densities can reveal which density regions are most relevant for specific observables in finite nuclei. 
For instance, the average density in nuclei is approximately 
$0.1\ \text{fm}^{-3}$~\cite{PhysRevLett.102.122502}, which is one reason why EoS parameters are often best constrained around this value. 


For each density $\rho$, the 100 Skyrme functionals provide 100 values of $K(\rho)$ and the corresponding values of $R_2$ and $\RR$. 
If variations in $K(\rho)$ systematically accompany variations in the density moments, then the corresponding quantities are correlated. We quantify the strength of this correlation among all three quantities using the multiple correlation coefficient (MCC)~\cite{allison1998multiple} 
\begin{equation}
       r_{YZ,X}^{2} = \frac{r_{YZ}^2+r_{ZX}^2-2r_{YZ}r_{YX}r_{ZX}}{1-r^2_{YZ}}.
\end{equation}
Similar to the PCC for two quantities, the MCC quantifies how well 
a quantity $X$
can be approximated by a linear combination of multiple variables -- in this case $Y=R_2$ and $Z=\RR$.
In this study, we primarily focus on the curvature of the energy per nucleon in symmetric nuclear matter at a given density,
i.e.,
$X=K(\rho)$, but the same approach can be applied to other EoS parameters as well.


\begin{figure}[h]
    \begin{center}
    \includegraphics[width=7cm]{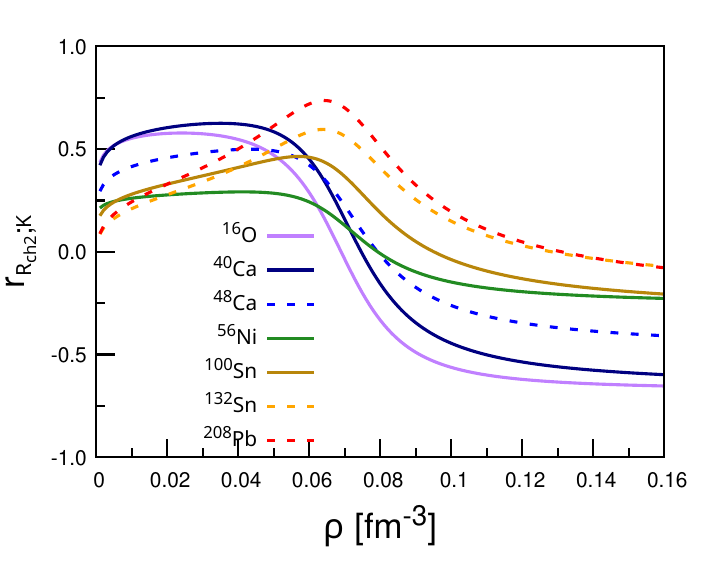}
    \includegraphics[width=7cm]{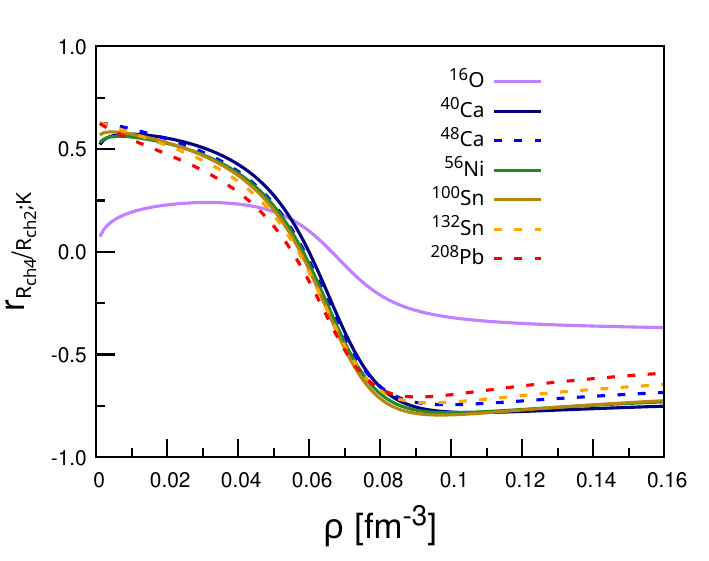}
    \includegraphics[width=7cm]{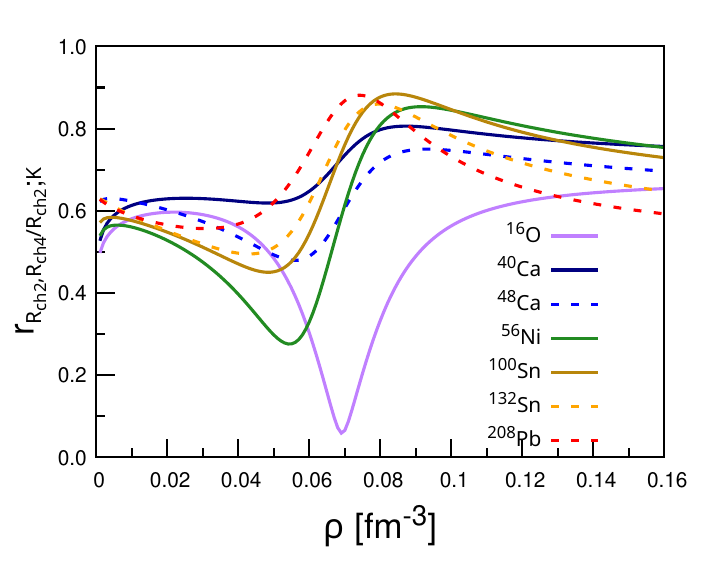}
    \end{center}
    \caption{Pearson correlation coefficients (PCC) between EoS parameters $K(\rho)$ and $R_2$ or $\RR$ and the MCC among all three for selected isospin-symmetric (solid lines) and asymmetric (dotted lines) nuclei.}
     \label{fig:3}
\end{figure}

At each density $\rho$,
we construct 100 sets of $\left(X(\rho), R_2, \RR\right)$
 using the 100 Skyrme functionals adopted in this study. From these data, we calculate both the PCC and the MCC.
This procedure is repeated for a set of representative nuclei, including $\ce{^16O}$, $\ce{^40Ca}$,
$\ce{^48Ca}$, $\ce{^56Ni}$, $\ce{^100Sn}$,
$\ce{^132Sn}$ and $\ce{^208Pb}$.

%
%
The resulting Pearson and multiple correlation coefficients (PCC and MCC)
of  $X=K(\rho)$
across various nuclei
are shown in 
Fig.~\ref{fig:3}.
%
%
The first two panels show the PCCs
between 
$K(\rho)$ and $R_2$,
and between $K(\rho)$ and $\RR$, respectively.
At low densities, $K(\rho)$
exhibits a positive correlation with both
$R_2$ and $\RR$.
However, as $\rho$ increases, the correlation gradually becomes negative. 
Notably, the PCCs with $\RR$ reveal a stronger anti-correlation at saturation density $\rho=\rho_0$.
As shown in the figure, the correlation changes sign
--- shifting from positive to negative ---
in the density range of approximately
$(0.06\sim 0.08)\ \text{fm}^{-3}$.
This behavior is consistently observed across all nuclei considered.
%
The accompanying positive correlation between $K(\rho)$ at low densities
is not surprising given that, at least for Skyrme models, $K(\rho)$ at low and at high densities are anti-correlated as demonstrated in Fig.~\ref{fig:1}.
A stronger anticorrelation is observed with $\RR$ than with $R_2$, and consistently for both symmetric and asymmetric nuclei, thus corroborating the connection between the compressibility and the surface diffuseness.
Only the light nucleus $^{16}\textmd{O}$ shows sizable quantitative deviations from this picture.
The MCCs, which account for correlations with both $R_2$ and $\RR$,
exhibit even stronger overall correlations,
as shown in the last panel of the figure.
%
In particular, the MCC for all 
the considered nuclei except $^{16}\textmd{O}$
is strongest at around $0.07-0.09$ fm$^{-3}$, {\em i.e.,} near half the saturation density. 
This density range corresponds to the surface region of the charge density distribution and is therefore expected to be most relevant for observables sensitive to surface diffuseness.
The correlations at densities of $0.1\textmd{fm}^{-3}$, which is close to the average density in heavy nuclei~\cite{PhysRevLett.102.122502}, 
are found generally weaker.
This is consistent with our argument that $K(\rho )$ is related to the surface of the density distribution (see also the discussion of Fig.~\ref{fig:2}) rather than the entire nucleus. 
%
%
The correlations of $K(\rho_0/2)$ with the moments of the density, where $\rho_0$ is the saturation density for each Skyrme model, were also checked. Although $\rho_0/2$ is
numerically 
quite close to $0.08$ fm$^{-3}$ for most
Skyrme models, we found that the correlation is noticeably weakened,
which means that the model-dependence on $\rho_0$
deteriorates the overall correlation.
The correlations at other fractions of $\rho_0$, including $K(\rho_0)$, were weaker than what we found at approximately $0.08$ fm$^{-3}$. 



\begin{figure}[h]
    \begin{center}
    \includegraphics[width=7cm]{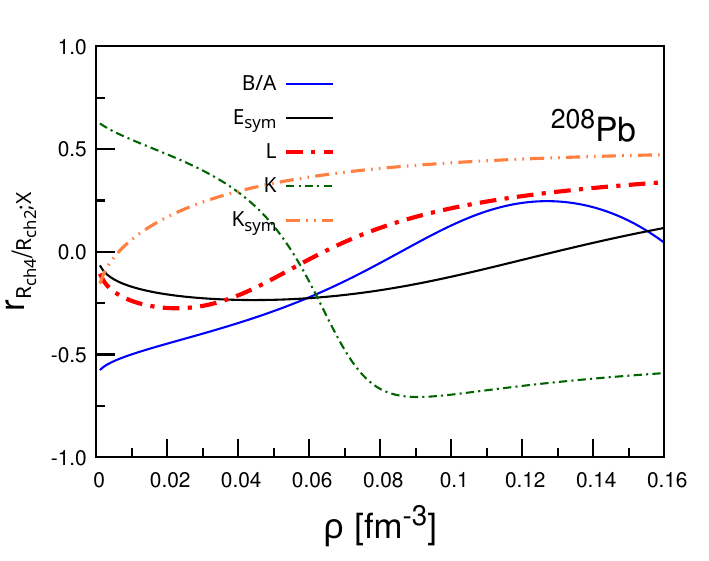}
    \includegraphics[width=7cm]{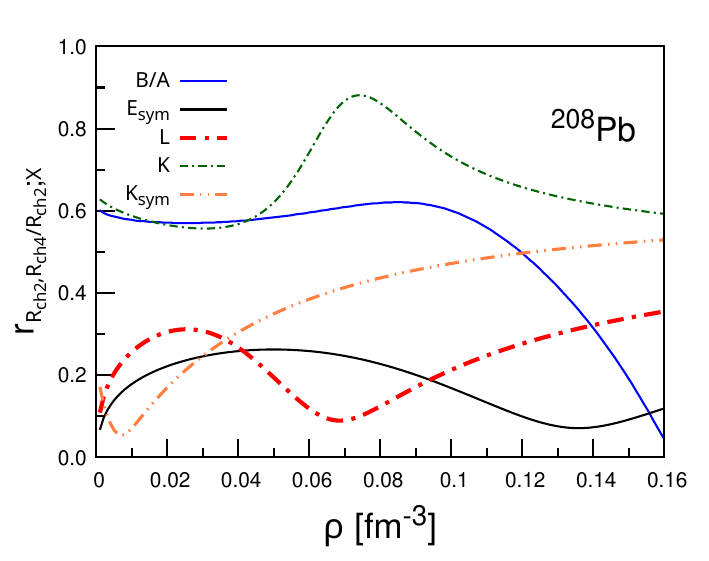}
    \end{center}
    \caption{Pearson correlation coefficients (PCC) and MCC results of various EoS parameters, $X(\rho)$, with $\RR$ and $R_2$ for $^{208}$Pb, as a function of density.}
     \label{fig:4}
\end{figure}

Let us also establish the relevance of $K(\rho)$ in comparison with the other EoS parameters.
Figure~\ref{fig:4} illustrates
the PCCs
of
various parameters
with $\RR$ (left panel) 
and the MCCs
(right panel)
as functions of 
the nuclear density $\rho$ in the case of $^{208}$Pb.
We observe that $\RR$ is generally less correlated with other EoS parameters than with $K(\rho)$.
Note that these trends hold for the moderately neutron-rich nuclei examined here, but not for highly exotic nuclei such as $^{60}$Ca, where we found that $L$ shows stronger correlations with $\RR$  than does the $K(\rho)$ of symmetric matter.


The trends presented above highlight 
the importance of focusing on specific density regions when examining the influence of EoS parameters on the density distributions, such as $\rho \simeq 0.08\,\text{fm}^{-3}$ in the case of $K(\rho)$.
%
%
Building on the observed correlations between $K(\rho)$ and the density moments $R_2$ and $\RR$, we next employ multi-parameter regression to obtain quantitative estimates of the curvature and to assess the extent to which these nuclear observables can constrain it.

\section{Constraining $K(\rho)$ from density moments}\label{sec:mpr}

Having identified 
\begin{equation}
X \equiv K(\rho),
\end{equation}
as the EoS quantity most strongly correlated with the density moments, we next ask whether $X$  can be estimated from measured values of $R_2$ and $\RR$ and with what precision.

To answer the above question,
we construct an estimator for $X$
as a function of $R_2$ and $\RR$
of a nucleus $\nu$,
for which 
we make use of
a standard multiple linear regression.
As detailed in the following subsections,
the fitting is carried out with the
Skyrme model predictions for $X$ as well as
nuclear moments of selected nuclei.
The fitting procedure and the resulting linear estimator are straightforward generalizations of ordinary linear regression (see, e.g., Ref.~\cite{Barlow1999}).
We then present a detailed analysis for $^{48}$Ca,
which is followed by our predictions
and the estimated uncertainties
for a few selected nuclei.
A test of the constructed model 
is presented in the final subsection.

\subsection{Linear estimator}
We begin by constructing
a linear estimator based on the values of $R_2$ and $\RR$. 
Let $X(i)$ denote the value of $X$ predicted by the $i$th Skyrme functional. 
Similarly,
$R_2^{(\nu)}(i)$ and $R_{4/2}^{(\nu)}(i)$ denote the
moments predicted by the same functional for nucleus $\nu$.
Assuming a linear dependence on the nuclear moments, we write
\begin{equation}
X(i)\simeq 
\alpha_X^{(\nu)} + \beta_X^{(\nu)} R_2^{(\nu)}(i) + \gamma_X^{(\nu)} R_{4/2}^{(\nu)}(i),
\label{eq:19}
\end{equation}
where the coefficients
$\left(\alpha_X^{(\nu)},\, \beta_X^{(\nu)},\, \gamma_X^{(\nu)}\right)$
are
the nucleus-specific coefficients to be 
determined.
We determine these coefficients through a multivariate fit using the data sets $\left\{X(i),\, R_2^{(\nu)}(i),\, \RR^{(\nu)}(i)\right\}$
obtained from the Skyrme energy density functionals adopted in this work.
The superscript $\nu$
labels the nucleus,
whereas
the index $i=1,2,\cdots,100$ labels the Skyrme functionals. 
The fit is carried out using the routines of the GNU Scientific Library \cite{gsl2021} under the header of $\texttt{gsl\_multifit\_linear}$.
Once the coefficients for a nucleus $\nu$ are given,
then the estimator enables us
to predict $X$ with the measured moments of 
the nucleus $\nu$,
\begin{equation}
X\simeq 
\alpha_X^{(\nu)} + \beta_X^{(\nu)} R_2^{(\nu)} + \gamma_X^{(\nu)} R_{4/2}^{(\nu)}
\equiv \tilde{X}^{(\nu)}\left(R_2^{(\nu)},\RR^{(\nu)}\right),
\label{eq:20}
\end{equation}
where we have defined our estimator
$\tilde{X}^{(\nu)}$.
The above equation becomes exact only when
the following two conditions
are satisfied. The first condition
is that the correlation of $X$ with the nuclear moments
is strong enough to the degree that it can be 
regarded as a function of $R_2$ and $\RR$ 
of the considered nuclei,
and the second is that the functional dependence is
well captured by the linear terms in $R_2$ and $\RR$.
We will come back to this point at the end of this Section.

\subsection{Uncertainty estimation}
Let us assume that for a specific nucleus $\nu$ an experimental value for the second moment $R_2^{(\nu,\mathrm{exp})}$ is available with high precision and a subsequent independent measurement yields the fourth moment and the ratio $\RR^{(\nu,\mathrm{exp})}$ with experimental uncertainty 
$\V(\RR^{(\nu,\mathrm{exp})})$. 
The output results of the fit for this nucleus $\nu$, and Eq.~(\ref{eq:20}) can be used to extract the value of the EoS parameter $\tilde{X}$. The uncertainty in $\tilde{X}$ can be obtained as the square root of the variance, which is given by 
\begin{equation}
    \V(\tilde{X}) = Y\,
    C\,
    Y^T + \gamma_X^2 
     \V(\RR^{(\nu,\mathrm{exp})}) , 
\label{VarXtilde}
\end{equation}
where $Y=(1, R_2^{(\nu,\mathrm{exp})}, \RR^{(\nu,\mathrm{exp})})$ and $C$ is the covariance matrix of $(\alpha_X^{(\nu)},\,\beta_X^{(\nu)},\,\gamma_X^{(\nu)})$. Here, $\V(\cdot)$ denotes the statistical variance of the corresponding quantity. 
In the case of experimentally determined quantities, $\V(\cdot)$ is understood to correspond to the standard deviation squared. 
%
In the above error propagation, we omit a term 
$\beta_X^2 \V(R_2^{(\nu,\mathrm{exp})})$ 
because
the experimental uncertainty in $R_2^{(\nu,\mathrm{exp})}$ is expected to be much smaller than that in $\RR^{(\nu,\mathrm{exp})}$. We verify this with an example below.

In what follows, we will omit the index $\nu$ and explicitly specify each nucleus for simplicity. 
We will also omit the superscript $\mathrm{exp}$ with the understanding that the symbols $R_2$ and $\RR$ will represent experimental values. Finally, we will omit the subscript $X$ from the fit coefficients. 

\subsection{Numerical example}
Before reporting our results, let us go through the analysis of the $K(0.08$fm$^{-3})$ fit to the results for $^{48}$Ca as an example. 
The parameter values for $^{48}$Ca are found to be 
%
$\alpha = 2006.29 \, \text{MeV}$,
$\beta = 43.40 \, \text{MeV/fm}$
and $\gamma = -1912.85 \, \text{MeV}$.
The associated covariance matrix $C$
is given by 
\[
\begin{pmatrix}
50108.49 & 732.32~\mathrm{fm}^{-1}  & -48542.81 \\
732.32~\mathrm{fm}^{-1} & 127.93~\mathrm{fm}^{-2} & -1088.45~\mathrm{fm}^{-1} \\
 -48542.81 & -1088.45~\mathrm{fm}^{-1} & 48251.49
\end{pmatrix} ~\mathrm{MeV}^2 .
\]
The diagonal elements are the variances of the three variables, whose square roots yield the corresponding uncertainties 
$224$~MeV, $11.3$~MeV/fm and $220$~MeV
for $\alpha$, $\beta$ and $\gamma$,
respectively.
The off-diagonal elements indicate 
substantial
correlations among the fitted coefficients; 
$r_{\alpha\beta}=\frac{C_{\alpha\beta}}{\sqrt{C_{\alpha\alpha}C_{\beta\beta}}}= 0.289$,
$r_{\alpha\gamma}=-0.987$, and
$r_{\beta\gamma}=-0.438$.
Thus, the strongest correlation is the nearly perfect
anticorrelation between $\alpha$ and $\gamma$.
These correlations are fully taken into account in Eq.~(\ref{VarXtilde}).


Given that the experimental uncertainty in $R_2$ is $0.002$~fm, we find that its contribution to $\V(K(0.08$ fm$^{-3}))$, $\beta^2\sigma_{R_2}^2$ equals $0.0075$ MeV$^2$. For $\gamma^2\V (\RR)$ to be comparable to that value, we should have $\V(\RR)\approx 10^{-9}$, which is unrealistically small. We were therefore justified to neglect the contribution of $\V(R_2)$ in $\V (\tilde{K})$. 


Suppose, as an illustration of the relevance of $\RR$, that the fourth moment of the charge density could be measured and yielded the ratio $\RR = 1.080\pm 0.005$. Application of the extracted coefficients and variables would give $\tilde{K}(0.08~\mathrm{fm}^{-3}) = 91 \pm 10 $~MeV. If, on the other hand, the measured value was $\RR = 1.110\pm 0.005$, we would get $\tilde{K}(0.08~\mathrm{fm}^{-3}) = 34\pm 11 $~MeV. 

We note that an existing Fourier-Bessel analysis of the charge form factor of $^{48}$Ca~\cite{deVries1987,Liu:2021ofe} gives a ratio ${\RR}_{\mathrm{ch}} \approx 1.082$ without reporting an uncertainty.  Assuming an uncertainty of $0.01$, we obtain $\tilde{K}(0.08~\mathrm{fm}^{-3}) = 87 \pm 19 $~MeV.
Better precision would yield more meaningful constraints. For example, assuming an uncertainty of $0.005$ we obtain $\tilde{K}(0.08~\mathrm{fm}^{-3}) = 87 \pm 10 $~MeV. Below we present detailed results for four stable nuclei, for which precise measurements of  $R_4$ are a realistic prospect. 

\subsection{Results for selected nuclei}
\begin{figure*}[h]
    \begin{center}
    \includegraphics[width=14cm]{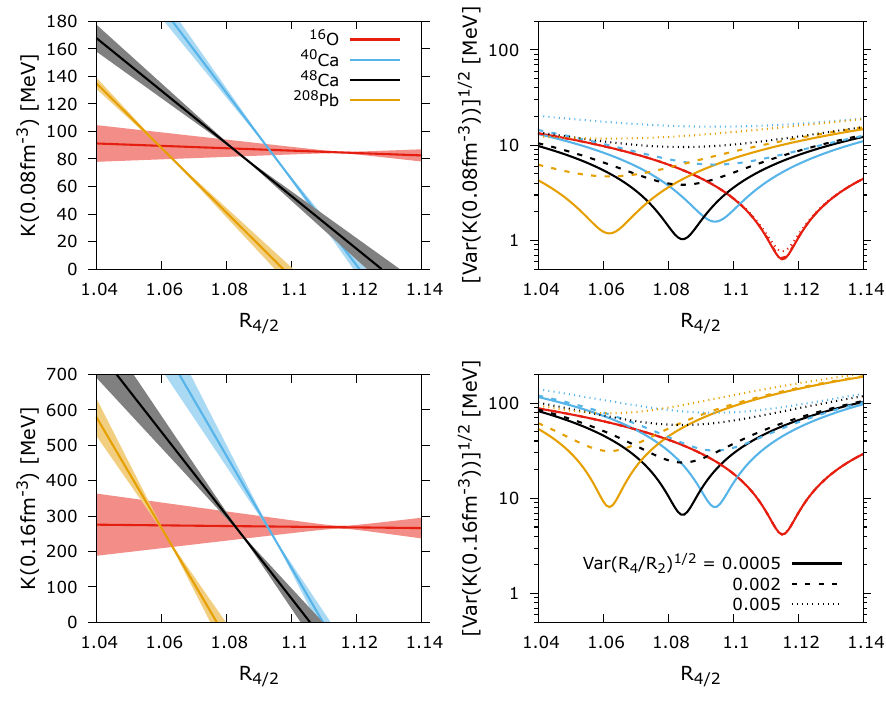}
    \end{center}
    \caption{%
    Left: Prediction and uncertainty for the curvature of the EoS at the indicated densities assuming accurate knowledge of $\RR$ as well as $R_2$ for the indicated nuclei, according to the analysis of one hundred Skyrme models. Right: The uncertainties in logarithmic scale assuming, in addition, an uncertainty of 0.0005 (full lines), 0.002 (dashed lines), or 0.005 (dotted lines) for $R_4/R_2$.
     \label{fig:5}}
\end{figure*}

In Fig.~\ref{fig:5}, left, we plot $\tilde{K}(0.08\;\textmd{fm}^{-3})$ and $\tilde{K}(0.16\;\textmd{fm}^{-3})$ with their uncertainties, as predicted for supposedly known $\RR$ and without the contribution of $\gamma^2 \V(\RR)$. The uncertainty bands come from the fits' covariance matrices, which is the first term of Eq.~(\ref{VarXtilde}).
On the right, we plot the corresponding uncertainties, Eq.~(\ref{VarXtilde}), by including the contribution of $\gamma^2 \V(\RR)$ to $\V{(K)}$ and 
by assuming the representative values of $\sqrt{\V(\RR)} = 0.0005$, $0.002$, and $0.005$ for the experimental uncertainty in $\RR$.
%
Although we plot the results only over the range of $\RR$ values allowed by the Helm model (see Sec.~\ref{RR_and_surface} for details), they are valid for all possible density profiles predicted by the Skyrme EDF models and 
are independent of the Helm-model density profile.
We note that all Skyrme models for all four nuclei predict values of $\RR$ well within this range. 
Next, with the help of these results, we assess the validity of the analysis and search for cases where a measurement of $R_4$ would provide meaningful constraints.

First, we observe that the value of $K(0.08 \; \textmd{fm}^{-3})$ determined from $^{16}$O shows a very mild dependence on $\RR$. For the range going from the homogeneous sphere case to the Gaussian density profile, {\em i.e.}, the range allowed by the Helm model, it restricts $K(0.08 \; \textmd{fm}^{-3})$ to lie between about 78 and 104 MeV. The results for this nucleus seem to favor a high $K(0.16 \; \textmd{fm}^{-3})$, but with large uncertainties at low and high $\RR$. It does not seem to be a good candidate for constraining $K(\rho)$ via a measurement of $\RR$. The reason might be simply that $^{16}$O is generally not a heavy enough nucleus to be reliably described with the Skyrme model.

Except for low values of $K(0.08 \; \textmd{fm}^{-3})$, below about $40$~MeV, the ratio $\RR$ is predicted larger for $^{40}$Ca than for $^{48}$Ca. In other words, the surface of the proton density distribution of $^{40}$Ca is predicted more diffuse than that of $^{48}$Ca. Experimentally measured charge densities also suggest a thicker surface for $^{40}$Ca than for $^{48}$Ca---see, {\em e.g.}, \cite{Friedrich:1982esq} and references therein for empirical estimates in these and other nuclei. Thus, our analysis is consistent with existing data.


Next, we turn to the values and uncertainties for the curvature $\tilde{K}(0.08\;\textmd{fm}^{-3})$ and $\tilde{K}(0.16\;\textmd{fm}^{-3})$ shown on the right of Fig.~\ref{fig:5}, corresponding to the representative values $\sqrt{\V(\RR)}=0.0005,0.002$, and $0.005$. 
The figures show the level of precision that would be reached for $\tilde{K}(0.08\;\textmd{fm}^{-3})$ and $\tilde{K}(0.16\;\textmd{fm}^{-3})$ given a measurement of a specific value of $\RR$ with a given uncertainty $\sqrt{\mathrm{Var}(\RR})$.
We notice that a measurement of $\RR$ for any of these nuclei, regardless of its value, could give a prediction for the curvature at $0.08\;\textmd{fm}^{-3}$  with an uncertainty of less than $20$~MeV. 
If the Skyrme model turns out to very accurately describe $\RR$, the precision could be much higher, as demonstrated by the dips at various values of $\RR$.
By contrast, the prediction that could be obtained for the curvature at $0.16\;\textmd{fm}^{-3}$ would carry an uncertainty almost one order of magnitude higher. 
Thus, a meaningful constraint on the curvature at the subsaturation density could be obtained if $R_4$,
or equivalently the ratio $\RR$, could be measured to a precision of 0.2\%, 0.5\%, or similar. 
In particular, if $R_4$ is measured within $0.5\%$ or better, the uncertainty expected for $K(0.08$~fm$^{-3})$ is at most $20$~MeV and can be as low as $1-2$~MeV depending on how accurate the Skyrme modeling turns out to be in describing the nuclear surface diffuseness.
Achieving the required sub-percent precision directly through measurements of the electric form factor at very low momentum transfer could be challenging. However, such precision may be attainable through combined analyses of experimental measurements and theoretical 
modeling~\cite{Xie2023}.

\subsection{A validation test}
\begin{figure}[h]
    \begin{center}
    \includegraphics[width=7cm]{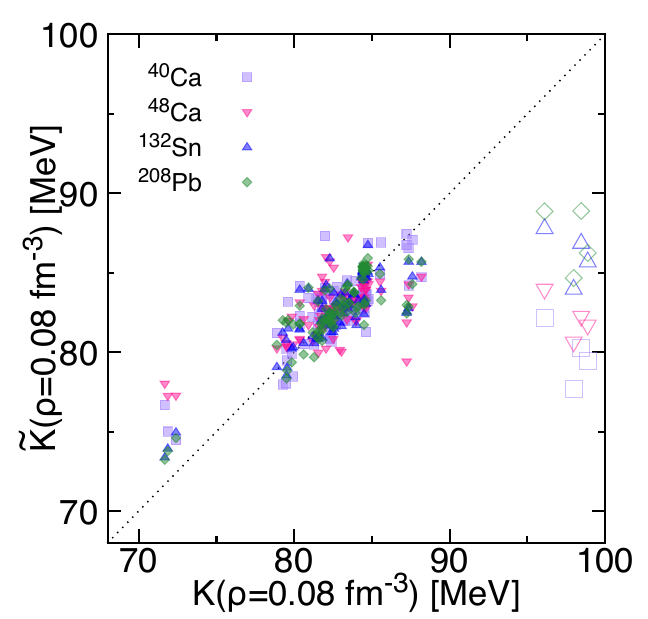}
    \end{center}
    \caption{The quantity $\tilde{K}(0.08\;\textmd{fm}^{-3})$ defined in Eq.~(\ref{eq:20}) for each functional and nucleus compared with the model curvature. The open symbols correspond to selected relativistic functionals for comparison.
    The size of each point is set proportional to
    $r/\sqrt{(\delta r)^2 +1 }$ with $\delta r= (r-r_\text{exp})/(0.01\ \text{fm})$,
    such that data points appear larger when the model prediction is closer to the experimental value.}
     \label{fig:6}
\end{figure}

Finally, Figure 6 serves as a validation test of the estimator defined in Eq. (\ref{eq:20}).
Specifically, the $\tilde{K}(0.08\;\textmd{fm}^{-3})$
obtained with Eq.~(\ref{eq:20})
will be identical to
${K}(0.08\;\textmd{fm}^{-3})$
if
the correlation of the incompressibility with $R_2$ and $\RR$ is perfect 
and the correlation can be 
represented as a linear function
of $R_2$ and $\RR$.
%
Thus
any deviation from the equality $\tilde{K}(0.08${fm}$^{-3}) = {K}(0.08\;\textmd{fm}^{-3})$ represented by the diagonal line
in Fig. 6
indicates a departure from the above assumptions. 
To see how well these assumptions work,
%
for four representative nuclei and the 100 Skyrme models, 
we compare the estimates 
$\widetilde K(0.08~\mathrm{fm}^{-3})$
obtained from Eq.~(\ref{eq:20}) with the values
$K(0.08~\mathrm{fm}^{-3})$ calculated directly from the
corresponding Skyrme functionals.
The results are shown in Fig.~\ref{fig:6} with full symbols. 
To examine whether our conclusions are specific to Skyrme functionals,
we have included results from four relativistic mean-field (RMF) models: 
NL3, NL3$^*$, NL-SH, and NL-RA1 \cite{Lalazissis:1996rd,Lalazissis:2009zz,Rashdan:2001nc}. 
These models, which include non-linear $\sigma$ meson terms, are among the standard RMF approaches~\cite{Dutra:2014qga} and have been successful in describing bulk properties of doubly magic nuclei.
We apply Eq.~(\ref{eq:20}) to each nucleus with the coefficients derived from the Skyrme models, but for $R_2$ and $\RR$, we insert the values obtained from each relativistic model. The results for the RMF models are shown in Fig.~\ref{fig:6} with open symbols.
%

%
%
In Figure~\ref {fig:6}, the size of each point reflects the accuracy of the corresponding model with respect to the empirical root-mean-square radius. Specifically, larger points represent smaller 
deviations of $R_2^{(\nu)}$ from the empirical 
value of $R_2^{(\nu,\mathrm{exp})}$. 
Note that the point size does not scale linearly with the deviation, in order to enhance visual clarity.

From Figure \ref{fig:6}, we observe a reasonable correlation between $K(\rho)$ and $\tilde K(\rho)$ at $\rho = 0.08\ \text{fm}^{-3}$, indicating that the incompressibility is well captured by a multivariable linear regression in terms of $R_2$ and $\RR$. 
The spread of the points is comparable to the uncertainties derived from the fits.
A certain degree of model dependence is evident, as the predictions from the four RMF approaches do not align with those from the Skyrme functionals. 
On the other hand, these four RMF models reproduce the empirical radii less accurately than the shown Skyrme models.

A more comprehensive analysis involving a broader set of RMF models would be desirable to further explore this behavior.

\section{Summary\label{sec:summary}}

This study explores the relationship between the nuclear EoS and the higher-order moments of the 
charge
density distribution, which depends on the surface diffuseness. Specifically, the study focuses 
on the curvature of 
the energy per particle, $K(\rho)$, of symmetric matter and
on the ratio 
$R_4/R_2$,
where 
$R_4$ and $R_2$ denote the fourth root of the fourth moment
and the root-mean-square radius
of the 
charge density,
respectively.
Utilizing one hundred Skyrme functionals and covariance analyses, this study investigates the correlation between $K(\rho)$ at various densities and the density moments ($R_2$, $\RR$) of symmetric and asymmetric nuclei. 


Among key findings is the identification of strong correlations among 
$K(\rho)$, $R_2$, 
$R_4/R_2$
at $\rho = 0.08 \, \text{fm}^{-3}$.
The analysis demonstrates that meaningful constraints
on the curvature at that density 
can
be obtained if $R_4$,
or equivalently the ratio $R_4/R_2$, 
is measured 
with
a precision of approximately
 $0.5\%$
in nuclei such as
$^{48}$Ca and $^{208}$Pb. 
In particular, if $R_4$ is measured within $0.5\%$ or better, the uncertainty expected for $K(0.08$~fm$^{-3})$ can be as low as $1-2$~MeV depending on how accurate the Skyrme modeling turns out to be in describing the nuclear surface diffuseness.



\section*{Acknowledgments}
This work 
was supported by the Institute for Basic Science (IBS-R031-D1), 
the National Research Foundation of Korea (NRF), funded by Ministry of Science and ICT (RS-2024-00436392) and by the Institute for Basic Science (IBS) through the NRF (2013M7A1A1075764).

\appendix
\section{Qualitative interpretation of the crossing of density profiles}
Here we provide an analytical illustration
of
why density profiles with different diffuseness cross at approximately $\rho(0)/2$
for given normalization and r.m.s. radius, utilizing
the Helm model
and the trapezoid distribution.


In the Helm model, 
the radial density is given as
(see Ref.~\cite{ANDRAE2000413}, Eq. (221)) 
\begin{eqnarray}
\rho(r)&=&
\frac{\rho_0}{2}
\left[
\operatorname{erf}
\!\left(
\frac{R_H+r}{R_G}
\right)
-
\operatorname{erf}
\!\left(
\frac{r-R_H}{R_G}
\right)
\right]
\nonumber\\
&&-
\frac{\rho_0\,R_G}{2\sqrt{\pi}\,r}
\left[
e^{-(r-R_H)^2/R_G^2}
-
e^{-(R_H+r)^2/R_G^2}
\right].
\label{eq:helm_exact}
\end{eqnarray}
%
Evaluation of Eq.~(\ref{eq:helm_exact}) 
around $r=R_H$ tells us that
the half-density radius $r_{1/2}$ is slightly smaller than
$R_H$,
\begin{equation}
\frac{r_{1/2}}{R_H}
=
1-
\frac{R_G^2}{2 R_H^2}+
\mathcal{O}
\!\left(
\frac{R_G^4}{R_H^4}
\right).
\end{equation}
For heavy nuclei 
the ratio $R_G/R_H$ becomes small, 
so the half-density radius is
well approximated by the
Helm radius $R_H$.


In the case of the trapezoid distribution, the analytical derivation is simpler than for the Helm model. 
Consider the trapezoidal density profile
\begin{equation}
\rho(r)=
\begin{cases}
\rho_0, & r\le c-\dfrac{t}{2},\\[1ex]
\rho_0\left(\frac12-\dfrac{r-c}{t}\right),
& c-\dfrac{t}{2}<r<c+\dfrac{t}{2},\\[2ex]
0, & r\ge c+\dfrac{t}{2},
\end{cases}
\end{equation}
where $c$ denotes the half-density radius and $t$ the surface thickness.
It can be easily verified that the half-density radius is 
\begin{equation}
r_{1/2} = c. 
\end{equation}

The normalization is
\begin{equation}
N
=
4\pi\int_0^\infty r^2\rho(r)\,dr
=
4\pi\rho_0
\left(
\frac{c^3}{3}
-\frac{ct^2}{12}
\right),
\end{equation}
while the mean-square radius is
\begin{equation}
\langle r^2\rangle
=
\frac{4\pi}{N}
\int_0^\infty\! r^4\rho(r)\,dr
=
\frac{3}{5}c^2
+
\frac{1}{20}t^2
+
\mathcal{O}\!\left(\frac{t^4}{c^2}\right).
\end{equation}
Thus, density profiles with the same rms radius imply
$\delta\langle r^2\rangle \simeq 
\frac{6}{5}c\,\delta c
+\frac{1}{10}t\,\delta t \simeq 0$,
or
\begin{equation}
\delta c
\simeq
-\frac{t}{12c}\,\delta t.
\end{equation}
Since $t/c\ll1$, the half-density radius is only weakly affected by changes in the surface thickness. For example, with
$c\sim 4~\mathrm{fm}$ and
$t\sim 2~\mathrm{fm}$,
one obtains
\begin{equation}
\delta c \approx -0.04\,\delta t.
\end{equation}

Thus, even substantial variations in the diffuseness produce only small shifts of the half-density radius. As a result, profiles with the same normalization and r.m.s. intersect in the vicinity of the half-density point.
This argument is consistent with the microscopic result observed in Fig. 2, namely that the crossing
occurs close to the half-density region for stable nuclei.

\bibliographystyle{elsarticle-num}
\bibliography{ref.bib}
\end{document}